\documentclass[]{aastex631}

\received{August 13, 2024}
\accepted{November 7, 2024}
\submitjournal{Astrophysical Journal Letters}

\usepackage{graphicx}
\usepackage[caption=false]{subfig}

\newcommand{\ha}{H$\alpha$}
\newcommand{\hb}{H$\beta$}

\newcommand{\lya}{Ly$\alpha$}

\shorttitle{Condor Discovery of the Cosmic Web Toward the M81 Group}
\shortauthors{Lanzetta et al.}

\begin{document}

\title{Introducing the Condor Array Telescope.  VI.  Discovery of Extensive 
Ionized Gaseous Filaments of the Cosmic Web in the Direction of the M81 Group}

\author{Kenneth M.\ Lanzetta}
\affiliation{Department of Physics and Astronomy,
Stony Brook University,
Stony Brook, NY 11794-3800, USA}

\author{Stefan Gromoll}
\affiliation{Amazon Web Services,
410 Terry Ave.\ N,
Seattle, WA 98109, USA}

\author{Michael M.\ Shara}
\affiliation{Department of Astrophysics,
American Museum of Natural History,
Central Park West at 79th St.,
New York, NY 10024-5192, USA}

\author{David Valls-Gabaud}
\affiliation{Observatoire de Paris,
LERMA, CNRS,
61 Avenue de l'Observatoire,
75014 Paris, FRANCE}

\author{Frederick M.\ Walter}
\affiliation{Department of Physics and Astronomy,
Stony Brook University,
Stony Brook, NY 11794-3800, USA}

\author{John K. Webb}
\affiliation{Institute of Astronomy,
University of Cambridge,
Madingley Road,
Cambridge CB3 9AL, UNITED KINGDOM}

\begin{abstract}
We used the Condor Array Telescope to obtain deep imaging observations through
luminance broad-band and He~II, [O~III], He~I, H$\alpha$, [N~II], and [S~II]
narrow-band filters of an extended region of the M81 Group spanning $\approx 8
\times 8$ deg$^2$ on the sky centered near M81 and M82.  Here we report aspects
of these observations that are specifically related to (1) a remarkable
filament known as the ``Ursa Major Arc'' that stretches $\approx 30$ deg on the
sky roughly in the direction of Ursa Major, (2) a ``Giant Shell of Ionized
Gas'' that stretches $\approx 0.8$ deg on the sky located $\approx 0.6$ deg NW
of M82, and (3) a remarkable network of ionized gaseous filaments revealed by
the new Condor observations that appear to connect the arc, the shell, and
various of the galaxies of the M81 Group and, by extension, the group itself.
We measure flux ratios between the various ions to help to distinguish
photoionized from shock-ionized gas, and we find that the flux ratios of the
arc and shell are not indicative of shock ionization.  This provides strong
evidence against a previous interpretation of the arc as an interstellar shock
produced by an unrecognized supernova.  We suggest that all of these objects,
including the arc, are associated with the M81 Group and are located at roughly
the distance $\approx 3.6$ Mpc of M81, that the arc is an intergalactic
filament, and that the objects are associated with the low-redshift cosmic web.
\end{abstract}


\section{Introduction}

Over the past several years, we have used the Condor Array Telescope
\citep{lan2023a} to obtain deep imaging observations through luminance
broad-band and He~II 468.6 nm, [O~III] 500.7 nm, He~I 587.5 nm, H$\alpha$,
[N~II] 658.4 nm, and [S~II] 671.6 nm narrow-band filters of an extended region
of the M81 Group comprising 13 adjacent ``Condor fields''\footnote{``Condor
fields'' are a set of fields with field centers that tile the entire sky with
the Condor field of view, allowing for overlap.} spanning $\approx 8 \times 8$
deg$^2$ on the sky centered near M81 and M82.  Details of some of these
observations are reported in a companion paper (\citealp{lan2024a}, hereafter
Paper V).  These new Condor observations reveal extensive extended structures
of ionized gas in the direction of the M81 Group, some of which are plausibly
located within the Galaxy and some of which are plausibly located at the
distance of M81.

Here we report aspects of these observations that are specifically related to
(1) a remarkable filament known as the ``Ursa Major Arc'' \citep{mcc2001,
bra2020a} that stretches $\approx 30$ deg on the sky roughly in the direction
of Ursa Major, (2) a ``Giant Shell of Ionized Gas'' \citep{lok2022} that
stretches $\approx 0.8$ deg on the sky located $\approx 0.6$ deg NW of M82, and
(3) a remarkable network of ionized gaseous filaments revealed by the new
Condor observations that appear to connect the Ursa Major Arc, the Giant Shell
of Ionized Gas, and various of the galaxies of the M81 Group and, by extension,
the group itself.  We suggest that this is a direct-imaging observation of the
low-redshift cosmic web.

\section{Background and Context}

\subsection{Ursa Major Arc}

Over two decades ago, \citet{mcc2001} reported the discovery of faint \ha\
emission from a long, straight, narrow filament extending $\approx 2.5$ deg on
the sky in the direction of Ursa Major.  More recently, \citet{bra2020a}
reported observations obtained with GALEX \citep{mar2005} that show ultraviolet
emission from this filament extending $\approx 30$ deg on the sky, with a width
ranging from less than one up to a few arcmin.  This ``Ursa Major Arc'' has
been interpreted variously as an interstellar trail of ionized gas produced by
an unseen ionizing source \citep{mcc2001} and as an interstellar shock produced
by an unrecognized supernova \citep{bra2020a}.  The arc is clearly visible in
the \ha\ and [N~II] mosaic difference images of Paper V, stretching across (and
beyond) the entire images.  It is notable that the Ursa Major Arc passes within
only $\approx 45$ arcmin of the center of M81.

\subsection{Giant Shell of Ionized Gas}

Recently, \citet{lok2022} reported the discovery of a shell of ionized gas that
stretches $\approx 0.8$ deg on the sky located $\approx 0.6$ deg NW of M82
based on deep imaging observations through \ha\ and [N~II] 658.4 nm narrow-band
filters of the vicinity of M81 and M82 obtained with the Dragonfly Spectral
Line Mapper pathfinder.  The authors argued that this ``Giant Shell of Ionized
Gas''  is associated with M82 and the M81 Group rather than with the Galaxy.
This interpretation is complicated by the low recession velocity of the group,
which makes it difficult to unambiguously distinguish emission from the group
from emission from the Galaxy, even with spectroscopic observations in hand.
The authors nevertheless speculated that the shell might plausibly be ejected
from M82 as a result of tidal interactions or the starburst activity of the
galaxy or that the shell might plausibly be falling into M82 from the cosmic
web in a way that is perhaps related to the superwind of the galaxy.  The
shell is clearly visible in the \ha\ and [N~II] mosaic difference images of
Paper V, which reveal new details of the shell, as is discussed in Paper V and
below.

\section{Condor Observations and Data Processing}

The Condor observations of the M81 Group described in Paper V consist of deep
imaging observations through luminance broad-band and He~II, [O~III], He~I,
\ha, [N~II], and [S~II] narrow-band filters of a extended region of the M81
Group comprising 13 adjacent Condor fields spanning $\approx 8 \times 8$
deg$^2$ on the sky centered near M81 and M82.  The images were processed using
the Condor data pipeline \citep{lan2023, lan2023a}, and the various groups of
coadded images were combined into seven mosaic images---one each obtained
through the luminance and six narrow-band filters.

The M81 Group is located in a direction of significant Galactic cirrus, and
starlight scattered from the cirrus dominates the diffuse light seen in both
the broad- and narrow-band images.  To at least a first approximation, any line
emission is insignificant in comparison to continuum over the very wide
bandpass of the luminance filter, so the luminance image roughly traces
continuum while the narrow-band images trace line emission plus continuum.
Accordingly, the luminance mosaic image was subtracted from the narrow-band
mosaic images to yield difference mosaic images that trace more or less only
line emission, and regions of the resulting mosaic images around stars
contained in the Gaia DR3 catalog \citep{gai2017, gai2018, gai2021, gai2022}
were masked at an isophotal limit, replacing the values of the masked pixels
with the value of the median of nearby pixels.  Details of this procedure are
described in Paper V.

The net result of the observations and data processing described in Paper V is
six continuum-subtracted, masked mosaic images through the He~II, [O~III],
He~I, \ha, [N~II], and [S~II] narrow-band filters of an extended region
spanning $\approx 8 \times 8$ deg$^2$ on the sky centered near M81 and M82.

\section{Results}

Selected portions of the mosaic difference images through the \ha\ and [N~II]
filters are shown in Figure 1.  In Figure 1, the left panel shows the \ha\
image together with a schematic depiction of some of the features visible in
the images.  As is described in Paper V, this schematic depiction is not
intended to be exhaustive or definitive but rather is intended only to label
some of the possible features.  In Figure 1, the middle panel shows a portion
of the \ha\ image, and the right panel show a portion of the [N~II] image.  The
schematic depiction of the left panel of Figure 1 indicates known galaxies of
the M81 Group (shown in light green), apparent clouds of gas (shown in light
blue), filamentary or essentially one-dimensional structures (shown in pink),
and an apparent or possible bubble or shell (shown as open yellow circle).  M81
and M82 are visble near the left-hand edges of the panels (as galaxies A and B,
respectively), and NGC~2976 is visible below center of the panels (as galaxy
D).  We note the following:

\begin{figure}
\centering
\subfloat{
  \includegraphics[width=1.00\linewidth, angle=0]{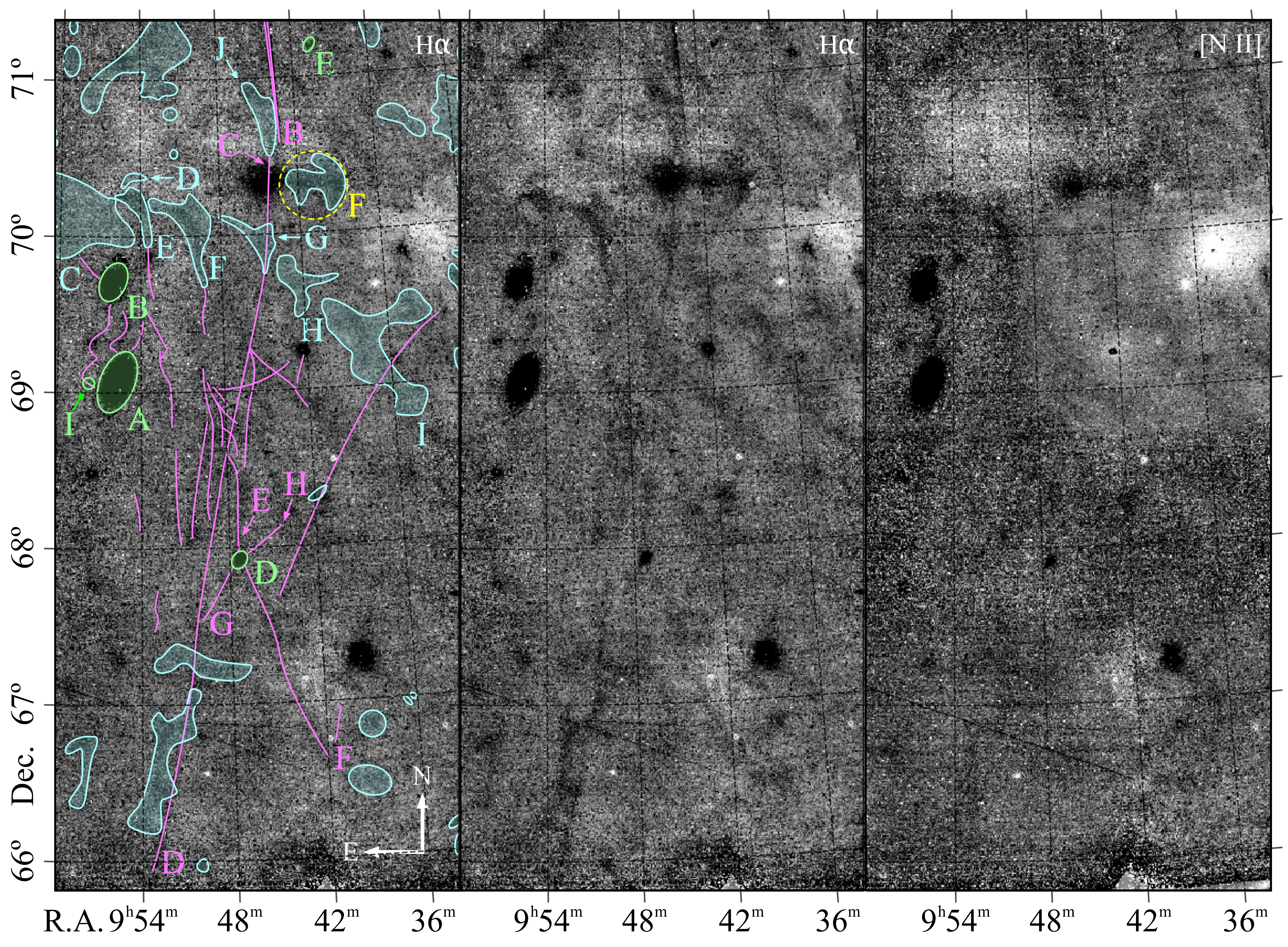}
}
\caption{Selected portions of mosaic difference images through \ha\ (left and
middle panels) and [N~II] (right panel) filters formed by subtracting luminance
broad-band image from narrow-band image and masking known stars.  Left panel
shows schematic depiction of some features visible in images.  Galaxies are
shown in light green, apparent clouds of gas are shown in light blue,
filamentary or essentially one-dimensional structures are shown in pink, and an
apparent or possible bubble or shell is shown as open yellow circle.  M81, M82,
and NGC~2976 are indicated in light green as galaxies A, B, and D,
respectively.  Clouds E and F are the Giant Shell of Ionized Gas, and filament
BCD is the Ursa Major Arc.  The bright unlabeled objects are stars.  Each image
spans $3.6 \times 5.5$ deg$^2$, and for each image N is up and E is to left.
Images are displayed block processed by $32 \times 32$ pixels, as described in
Paper V.}
\end{figure}

First, the Ursa Major Arc is visible in \ha\ and [N~II], stretching roughly
N-S across (and beyond) the extent of the images.  The character of the arc
changes across the images in the sense that it is thicker and brighter toward
the N and thinner and fainter toward the S.  The arc appears to exhibit a
discontinuity coincident with cloud J, and the southern segment of the arc
appears to originate or terminate on cloud J.

Next, there is a remarkable network of criss-crossed ionized gaseous filaments
that appear to intersect and overlap the Ursa Major Arc and various of the
galaxies of the M81 Group.  The filaments are visible in \ha, and some of the
filaments are visible in [N~II].  The widths of the filaments are typically
$\approx 1$ arcmin, i.e.\ comparable to the width of the Ursa Major Arc.  The
apparent epicenter of the network is located $\approx 0.9$ deg SW of M81.
Remarkably, there are four filaments that appear to originate or terminate on
NGC~2976: one at roughly 12 o'clock (designated as filament E) and others at
roughly 2 o'clock (filament H), 5 o'clock (filament F), and 7 o'clock
(filament G).  Filaments E and G appear to intersect the Ursa Major Arc toward
the E (east) of the galaxy.  Further, there is one filament that appears to
originate or terminate on NGC~3077 (which is off the left-hand edges of and so
not shown in the panels of Figure 1) at roughly 8 o'clock, and there are at
least three filaments that appear to originate or terminate on M82, including
filaments at roughly 2, 4, and 10 o'clock.  Expanded portions of the mosaic
difference image through the \ha\ filter centered on NGC~2976, NGC~3077, and
M82 are shown in Figure 2, in which these various filaments are visible.  It is
notable that the Ursa Major Arc passes within only $\approx 10$ arcmin of the
center of NGC~2976.  The middle and right panels of Figure 2 also show a
filament located $\approx 15$ arcmin N of NGC~3077 and M82 and running roughly
in the direction SE to NW that does not appear to originate or terminate on any
galaxy.

\begin{figure}
\centering
\subfloat{
  \includegraphics[width=0.30\linewidth, angle=0]{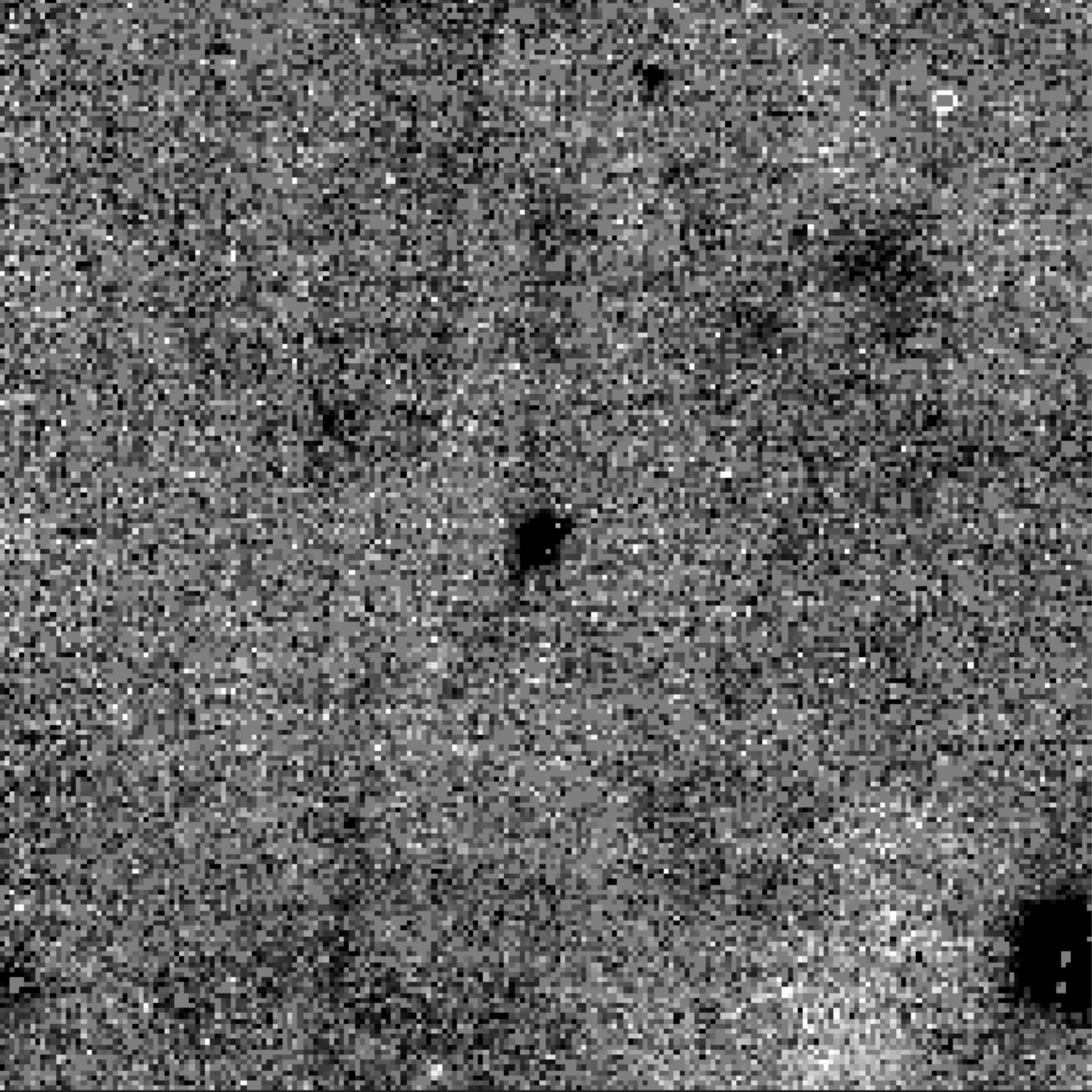}
}
\subfloat{
  \includegraphics[width=0.30\linewidth, angle=0]{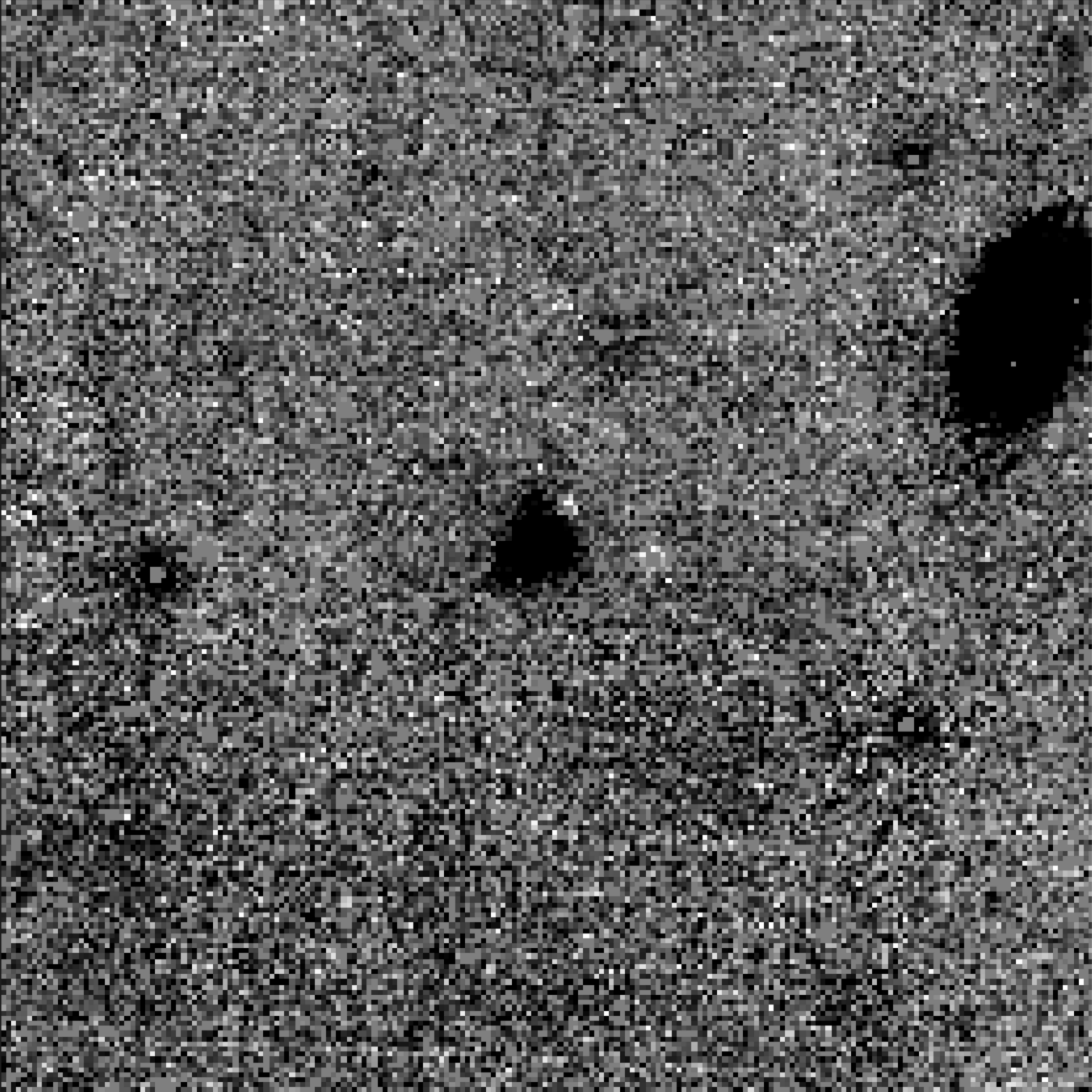}
}
\subfloat{
  \includegraphics[width=0.30\linewidth, angle=0]{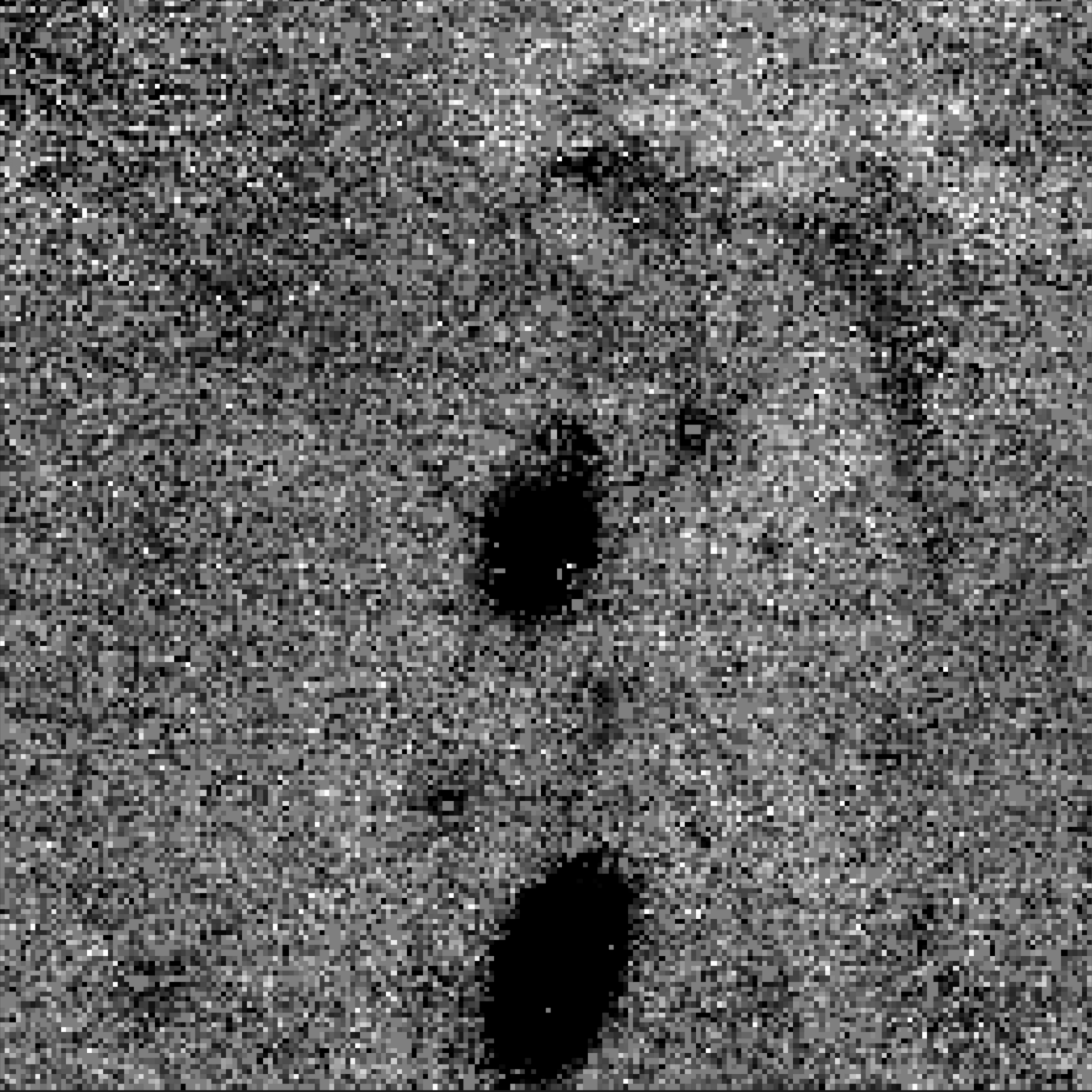}
}
\caption{Expanded portions of mosaic difference image through \ha\ filter
centered on NGC~2976 (left panel), NGC~3077 (middle panel), and M82 (right
panel).  Each image spans $1.6 \times 1.6$ deg$^2$, and for each image N is up
and E is to left.  Images are displayed block processed by $32 \times 32$
pixels, as described in Paper V.}
\end{figure}

Finally, the Giant Shell of Ionized Gas is visible in \ha\ and [N~II].  The
shell is clearly comprised of at least two pieces:  an eastern piece designated
as cloud E, and a western piece designated as cloud F.  Both pieces exhibit
similar morphologies, with a shape that resembles a ``comma.'' Further, both
pieces appear to exhibit thin filaments that extend toward the S.  The filament
of cloud E appears to connect to the filaments of the western side of M82, and
the filament of cloud F appears to connect to the network of criss-crossed
filaments further to the S.  It is notable that the Ursa Major Arc passes
within only $\approx 20$ arcmin of the western-most edge of cloud F.  There is
another cloud designated as cloud G located immediately W of cloud F that
exhibits a morphology similar to those of clouds E and F, with a shape that
resembles a comma; this cloud is exactly coincident with the Ursa Major Arc,
and we suggest that it comprises part of the shell.  In fact, clouds E, F, and
G appear to form part of a larger chain of clouds that runs roughly NE to SW
and includes clouds H and I to the W and other clouds (which are not shown in
Figure 1 but are shown in Figure 8 of Paper V) K, L, M, and C to the E, and we
suggest that all of these clouds might comprise part of the shell.  Another
smaller cloud D (which is shown in Figure 1) may also be associated with this
chain and with the shell.  Cloud J, which as noted above is coincident with the
apparent discontinuity of the Ursa Major Arc, exhibits a morphology similar to
those of clouds E, F, and G, with a shape that resembles a comma.  Indeed there
are dozens of other clouds visible in \ha\ distributed throughout the mosaic
image, especially toward the N and NW of M81, some (or all) of which may (or
may not) be associated with the shell.  (Again these other clouds are not shown
in Figure 1 but are shown in Figure 8 of Paper V.)  Some details of the shell
are visible in the right panel of Figure 2.  The bubble or shell F may be
associated with the shell or the arc (or both).

\section{Discussion}

The results of \S\ 4 appear to show a direct connection between the Ursa Major
Arc, the Giant Shell of Ionized Gas, the network of criss-crossed ionized
gaseous filaments, and various of the galaxies of the M81 Group, including most
dramatically NGC~2976.  This apparent connection seems to imply that (1) all of
these objects, including the Ursa Major Arc, are associated with the M81 Group
and are located at roughly the distance $\approx 3.6$ Mpc of M81
\citep{fre1994} and (2) the Ursa Major Arc is an intergalactic filament rather
than an interstellar trail of ioinized gas or an interstellar shock.  Here we
consider additional evidence in support of this proposition.

\subsection{Spatial Coincidences}

The suggestion by \citet{bra2020a} that the Ursa Major Arc is an interstellar
shock produced by an unrecognized supernova is based primarily on their finding
that the arc appears to be comprised of several piecewise-continuous arclets
that roughly lie along $\approx 1$ rad of the circumference of a circle of
radius $29.28 \pm 0.23$ deg centered on Galactic coordinates $(l, b) =
(107^\circ.7, 60^\circ.0$).  The authors also identified other arclets in the
vicinity that do not lie along this circle.  This is illustrated in Figure 3,
which is based on Figure 2 of \citet{bra2020a}.  Figure 3 shows a stereographic
projection of a region of the sky centered on the purported center of the
circle, where the arclets that appear to lie along the circumference of the
circle are indicated as feature A and the arclets that do not lie along the
circle are indicated as features B1, B2, and C; Figure 3 also indicates the
boundaries of Figure 1, M81, and the approximate location of the M81 Group.

\begin{figure}
\centering
\subfloat{
  \includegraphics[width=0.40\linewidth, angle=0]{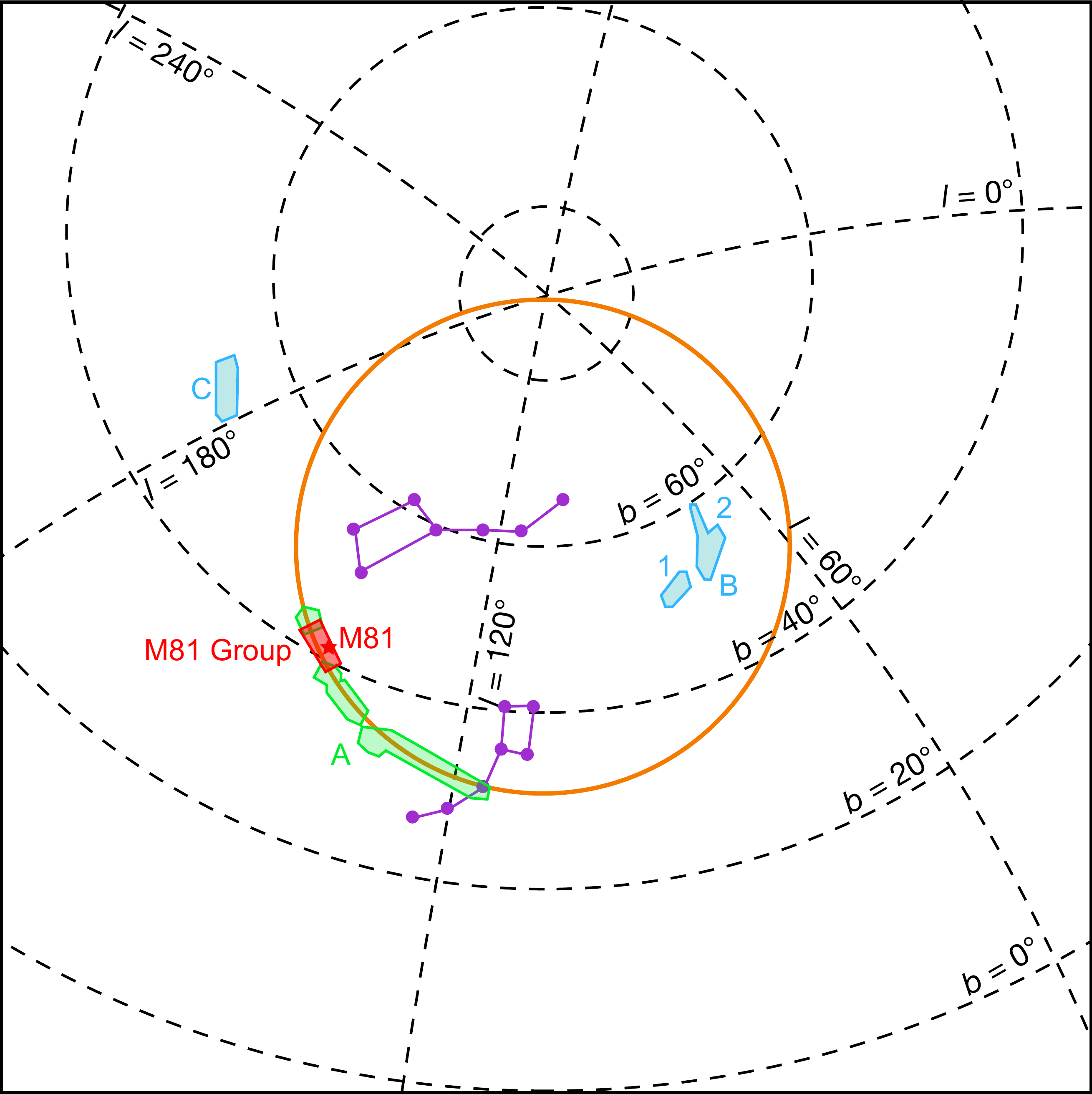}
}
\caption{Stereographic projection of region of sky centered on purported center
of circle along which arclets that comprise Ursa Major Arc appear to lie, based
on Figure 2 of \citet{bra2020a}.  Arclets that appear to lie along
circumference of circle are indicated as feature A and shown in green, and
arclets that do not lie along circle are indicated as features B1, B2, and C
and shown in blue. Boundaries of Figure 1 are shown as red polygon, M81 is
shown as red star, and approximate location of M81 Group is indicated.  Circle
of radius $29.28$ deg centered on Galactic coordinates $(l, b) = (107^\circ.7,
60^\circ.0$) is shown in orange.  Grid shows Galactic coordinates, and Ursa
Major and Ursa Minor are shown in purple.}
\end{figure}

It is clear from Figure 3 that there is a remarkable spatial coincidence
between the Ursa Major Arc and the M81 Group.  This sets into broader context
the previous statements that the Ursa Major Arc passes within $\approx 45$
arcmin of M81, $\approx 10$ arcmin of NGC~2976, and $\approx 20$ arcmin western
edge of cloud F and is exactly coincident with cloud G.  While it is indeed
striking that the arclets that comprise the Ursa Major Arc appear to roughly
lie along the circumference of a circle, it is perhaps no less striking that
the arc is essentially exactly aligned with one of the nearest galaxy groups
beyond the Local Group, which itself appears to be criss-crossed by ionized
gaseous filaments.  Further, it appears from Figure 3 that it is at least
plausible that, rather than tracing out the circumference of a circle, feature
A is actually part of a larger structure that extends toward increasing
Galactic latitude and longitude to encompass feature C (and perhaps even bends
toward decreasing longitude to encompass features B1 and B2).  We consider the
remarkable spatial coincidence between the M81 Group and the Ursa Major Arc to
provide strong circumstantial evidence in support of the proposition that the
arc is an intergalactic filament associated with the M81 Group.

\subsection{Ionization Mechanisms}

One of our primary motivations for acquiring the Condor observations of the M81
Group was to exploit the diagnostic capabilities of deep imaging observations
through [O~III], \ha, [N~II], and [S~II] narrow-band filters to determine or
constrain physical properties of ionized gas within the group.  Measurements of
flux ratios between these various ions can help to distiguish photoionized from
shock-ionized gas (e.g.\ \citealt{bal1981, vei1987}), which bears on the nature
and origin of the gas.  The previous imaging observations of the Ursa Major Arc
analyzed by \citet{bra2020a} were carried out in only \ha\ and continuum, and
the previous imaging observations of the Giant Shell of Ionized Gas analyzed by
\citet{lok2022} were carried out in only \ha, [N~II], and continuum
(supplemented by spectroscopic observations of a portion of the shell), which
limited the ability of these previous analyses to constrain the ionization
mechanisms at play.

We measured flux ratios between ions of selected features in the images by the
following procedure: First, we started with the $32 \times 32$ pix$^2$
block-processed mosaic images described in Paper V, because these images are
insensitive to faint point sources (e.g.\ intergalactic H II regions).  Next,
we defined polygonal regions bounding the selected features, and we summed the
energy fluxes of the various images within these regions to form the flux
ratios.  Next, we determined the median absolute deviations of the
pixel-by-pixel ratios of the images within the regions to form robust measures
of the spreads of the ratios.  We applied this procedure to several of the
features described in Paper V, including portions of (1) the Ursa Major Arc and
(2) the Giant Shell of Ionized Gas and for comparison (3) the ``\ha\ Cap,'' (4)
the ``Rattlesnake Head Nebula,'' and (5) the ``Giant \ha\ Bubble.''  The \ha\
Cap is a well-known feature in the vicinity of M82 \citep{dev1999, leh1999},
and the Rattlesnake Head Nebula and Giant \ha\ Bubble are features discovered
by the Condor observations presented in Paper V.  The measurement of the Ursa
Major Arc was made on the northernmost portion of the arc visible in Figure 8
of Paper V, where the arc is significantly brighter than it is over the extent
of Figure 1.  Two measurements of the Rattlesnake Head Nebula were made, one of
the SE portion and one of the NW portion.  The Condor observations cover \ha\
but not \hb, and so where necessary we adopted a value ${\rm H}\alpha / {\rm
H}\beta = 2.86$ appropriate for case-B photoionization \citep{ost1989}; typical
values of this ratio for shock-ionized gas are not too different from this
(e.g.\ \citealt{gas1984}), although this procedure does not account for
possible extinction due to dust.

Results of the analysis are presented in Figure 4.  Figure 4 shows the
measurements described above overlaid with results of the MAPPINGS III
``precursor plus shock'' models of \citet{all2008} spanning a range of shock
velocities, magnetic parameters, and abundances.  In Figure 4, the left and
right panels show $\log {\rm [O~III]}/{\rm H}\beta$ versus $\log {\rm
[N~II]}/{\rm H}\alpha$, the middle panel shows $\log {\rm [O~III]}/{\rm
H}\beta$ versus $\log {\rm [S~II]}/{\rm H}\alpha$, the left and middle panels
are overlaid with models of solar abundances, and the right panel is overlaid
with models of a range of abundances.  The two measurements of the Rattlesnake
Head Nebula are plotted separately in Figure 4, with portions to the SE and NW
indicated by a blue circle and square, respectively, and indicate that the flux
ratios vary significantly across the nebula.

\begin{figure}
\centering
\includegraphics[width=0.99\linewidth, angle=0]{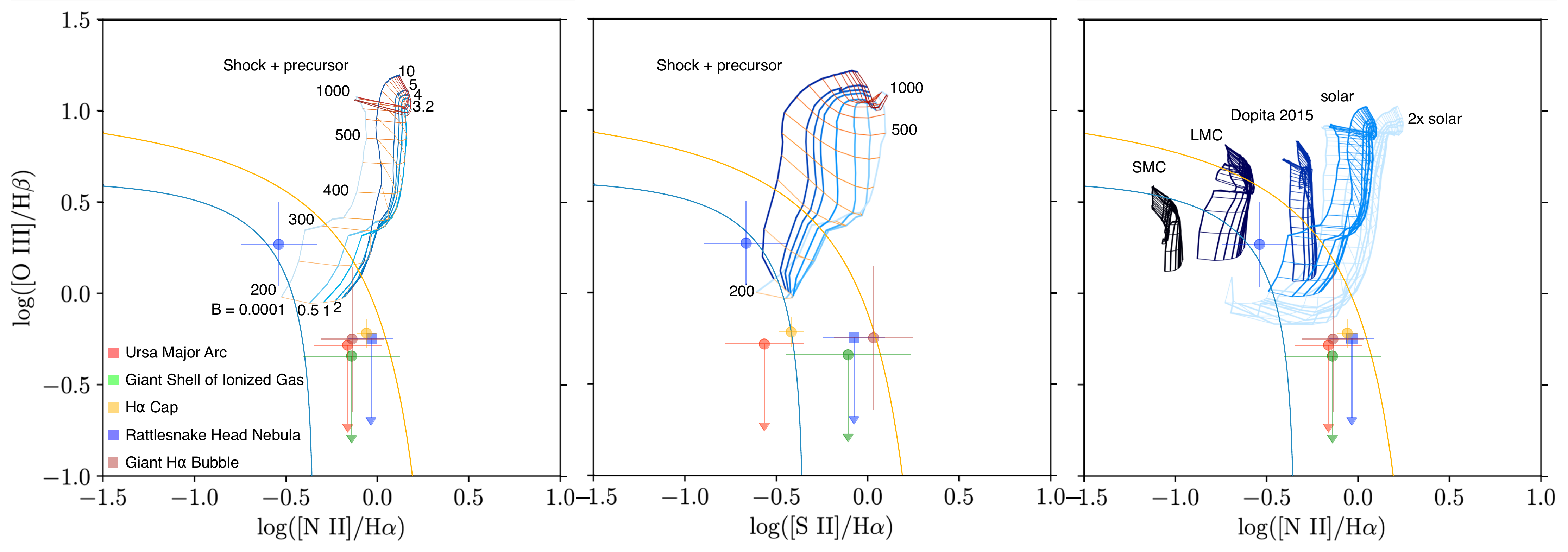}
\caption{Measured and modeled flux ratios between ions of selected features.
Measurements are of Ursa Major Arc (red), Giant Shell of Ionized Gas (green),
\ha\ Cap (orange), two portions of Rattlesnake Head Nebula (blue circle for SE
portion and square for NW portion), and Giant \ha\ Bubble (brown).  Models are
from MAPPINGS III ``precursor plus shock'' models of \citet{all2008} spanning a
range of shock velocities, magnetic parameters, and abundances (see their
figures 19, 21, and 23 for details).  LMC and SMC abundance patterns are
described by \citet{rus1992}.  Blue and orange curves show empirical fiducial
references set by \citet{bal1981} and \citet{kew2001}, respectively.  The flux
ratios provide strong evidence against the interpretation of the Ursa Major Arc
as an interstellar shock produced by an unrecognized supernova.}
\end{figure}

It is clear from Figure 4 that, except for the SE portion of the Rattlesnake
Head Nebula, the measured flux ratios of the various features are not
indicative of shock ionization but rather are indicative of photoionization.
In particular, the flux ratios of the Ursa Major Arc and the Giant Shell of
Ionized Gas are similar to each other and to those of the \ha\ Cap, the Giant
\ha\ Bubble, and the NW portion of the Rattlesnake Head Nebula.  This result
has two significant implications:  (1) It provides strong evidence against the
interpretation of the Ursa Major Arc as an interstellar shock produced by an
unrecognized supernova.  And (2) it suggests that the Ursa Major Arc and the
Giant Shell of Ionized Gas are similar and may share a common ionization
mechanism, which in turn bolsters the case that they share a common nature and
origin.  The measured flux ratios of the arc are also inconsistent with any of
the radiative-shock models considered by \citet{bra2020a} (which are based on
results of \citealt{sut2017}), which further undermines the interpretation as
an interstellar shock.  We consider the result that the flux ratios of the Ursa
Major Arc are not indicative of shock ionization to provide strong evidence in
support of the proposition that the arc is an intergalactic filament associated
with the M81 Group.

The possibility that the Giant Shell of Ionized Gas is photoionized by the
diffuse ultraviolet background radiation was considered in detail by
\citet{lok2022}, who found that the local background is insufficient to explain
the observed \ha\ surface brightness of the shell.  If the arc, the shell, and
the network of ionized filaments are indeed photoionized, then it is not clear
what is the source of the ionizing radiation.

\section{Conclusions}

Based on (1) the direct connection between the Ursa Major Arc, the Giant Shell
of Ionized Gas, the network of criss-crossed ionized gaseous filaments, and
various of the galaxies of the M81 Group, (2) the remarkable spatial
coincidence between the M81 Group and the Ursa Major Arc, and (3) the result
that the flux ratios of the Ursa Major Arc are not indicative of shock
ionization, we suggest that all of these objects are associated with the M81
Group and are located at roughly the distance $\approx 3.6$ Mpc of M81
\citep{fre1994}, that the Ursa Major Arc is an intergalactic filament, and that
objects are associated with the low-redshift cosmic web.  We suggest that this
is a direct-imaging observation of the low-redshift cosmic web.

If the Ursa Major Arc is indeed an intergalactic filament at the distance of
the M81 Group, then its angular extent of $\approx 30$ deg corresponds to a
length of $\approx 1.9$ Mpc, and its thickness, which ranges from $\approx 1$
to $10$ arcmin, corresponds to a width of $\approx 1$ to $10$ kpc.  This seems
thinner than might be expected for cosmic filaments but is consistent with the
thinest filaments seen in simulations of \lya\ emission from the high-redshift
cosmic web (e.g.\ \citealp{eli2020}), and perhaps only the
highest-surface-brightness portions of the filaments are visible.  It is
striking that the arc and the shell are of comparable surface brightness, with
both exhibiting a maximum surface brightness of $\approx 7 \times 10^{-18} \
{\rm erg} \ {\rm s} \ {\rm cm}^{-2} \ {\rm arcsec}^{-2}$ or ($\approx 1$ Ry)
and a typical surface brightness of $\approx 2.0 \times 10^{-18} \ {\rm  erg} \
{\rm s} \ {\rm cm}^{-2} \ {\rm arcsec}^{-2}$ (or $\approx 0.3$ Ry), at least
over the extent of the observations presented here.  Future efforts might seek
to analyze spatially-resolved images of the various flux ratios using
higher-signal-to-noise ratio images.

\begin{acknowledgments}
This material is based upon work supported by the National Science Foundation
under Grants 1910001, 2107954, 2108234, 2407763, and 2407764.  We gratefully
acknowledge the staff of Dark Sky New Mexico, including Diana Hensley and
Michael Hensley, for their superb logistical and technical support.  KML
acknowledges an enjoyable and productive visit to the Observatory of Paris,
where much of this manuscript was written.  The authors thank the anonymous
referee for very valuable comments.
\end{acknowledgments}

\software{
  astroalign \citep{ber2020},
  astropy \citep{2013A&A...558A..33A,2018AJ....156..123A},
  django \citep{django},
  Docker \citep{mer2014a},
  DrizzlePac \citep{gon2012},
  NoiseChisel \citep{akh2015, akh2019},
  numba \citep{lam2015},
  numpy \citep{har2020},
  photutils \citep{bra2020},
  scipy \citep{vir2020},
  SExtractor \citep{1996A&AS..117..393B}
}

\bibliography{manuscript}{}
\bibliographystyle{aasjournal}

\end{document}